\documentclass[conference]{IEEEtran}
\IEEEoverridecommandlockouts
\def\DRAFT
\usepackage{cite}
\usepackage{amsmath,amssymb,amsfonts}
\usepackage{algorithmic}
\usepackage[export]{adjustbox}
\usepackage{graphicx}
\usepackage{multirow}
\usepackage{textcomp}
\usepackage{xcolor}
\usepackage{hyperref}
\usepackage{multirow}
\usepackage{nicefrac}
\usepackage{balance}
\def\BibTeX{{\rm B\kern-.05em{\sc i\kern-.025em b}\kern-.08em
    T\kern-.1667em\lower.7ex\hbox{E}\kern-.125emX}}
\usepackage{bm}
\usepackage{mathtools}
\usepackage{amsmath}
\usepackage{multirow}
\usepackage{tabularx}
\usepackage{balance}
\usepackage{enumitem}
\usepackage{booktabs}
\usepackage[english]{babel}
\usepackage[autostyle,english=american]{csquotes}
\usepackage{graphics} %
\usepackage{threeparttable}
\usepackage{tikz}
\usetikzlibrary{automata,arrows,calc,positioning}

\ifx\DRAFT\fi

\begin{document}

\title{Tri-Spectral PPG: Robust Reflective Photoplethysmography by Fusing Multiple Wavelengths for Cardiac Monitoring
}

\author{\IEEEauthorblockN{Manuel Meier}
\IEEEauthorblockA{Department of Computer Science\\
ETH Z\"urich, Switzerland\\
manuel.meier@inf.ethz.ch}
\and
\IEEEauthorblockN{Berken Utku Demirel}
\IEEEauthorblockA{Department of Computer Science\\
ETH Z\"urich, Switzerland\\
berken.demirel@inf.ethz.ch}
\and
\IEEEauthorblockN{Christian Holz}
\IEEEauthorblockA{Department of Computer Science\\
ETH Z\"urich, Switzerland\\
christian.holz@inf.ethz.ch}
}

\maketitle

\begin{abstract}
Multi-channel photoplethysmography (PPG) sensors have found widespread adoption in wearable devices for monitoring cardiac health.
Channels thereby serve different functions---whereas green is commonly used for metrics such as heart rate and heart rate variability, red and infrared are commonly used for pulse oximetry.
In this paper, we introduce a novel method that simultaneously fuses multi-channel PPG signals into a single recovered PPG signal that can be input to further processing.
Via signal fusion, our learning-based method compensates for the artifacts that affect wavelengths to different extents, such as motion and ambient light changes.
We evaluate our method on a novel dataset of multi-channel PPG recordings and electrocardiogram recordings for reference from 10 participants over the course of 13 hours during real-world activities outside the laboratory.
Using the fusion PPG signal our method recovered, participants' heart rates can be calculated with a mean error of 4.5\,bpm (23\% lower than from green PPG signals at 5.9\,bpm).

\end{abstract}


\section{Introduction}

Reflective photoplethysmography (PPG) is a common sensing method to assess a wide range of physiological parameters, such as heart rate (HR), respiratory rate~\cite{daimiwal2014respiratory}, and peripheral oxygen saturation (SpO$_2$)~\cite{bagha2011real}.
Therefore, reflective PPG sensors are commonly built into devices such as fitness trackers (e.g., Fitbit, Garmin Fenix) or smartwatches (e.g., Apple Watch, Pixel Watch, Galaxy Watch) to monitor these parameters passively during wear.
PPG sensors obtain their signal from the variance of light absorption of arterial blood during the propagation of pulses.
Beyond the individual pulse as a whole, PPG sensors resolve the morphology of each blood volume pulse, including its various phases such as systolic and diastolic peaks in the form of small oscillations.

Robust PPG sensing is a challenging task, as PPG signals are subject to artifacts such as motion or optical interference~\cite{longmore_comparison_2019, wijshoff2011ppg}.
Reliable PPG analysis requires good signal quality;
robust signals facilitate individual pulse segmentation, such as for estimating a person's HR.
They also allow analyzing characteristics of a blood volume pulse through its morphology, such as to calculate temporal features for pulse transit time~\cite{li2014characters}.
PPG signals are thus commonly improved with established band-pass filters~\cite{ppg_filter} to aid feature extraction.

Depending on the magnitude of artifacts, however, conventional filtering can be insufficient for restoring the underlying blood volume pulses, as the signal's frequency spectrum typically overlaps with that of the artifacts.
Therefore, researchers have integrated accelerometers to detect motion~\cite{temko_accurate_2017} and remove artifacts via noise cancelation~\cite{asada2004active, yousefi2013motion, shimazaki2014cancellation}.
Alternatively, sensor designs have leveraged multiple light sources~\cite{abdallah_adaptive_2011} or photodiodes~\cite{vetter2010frequency} to harden PPG signal acquisition.
Multi-LED designs have become commonplace in today's commercial wearables (e.g., Garmin~\cite{garmin_garmin_nodate}), although their processing is proprietary.
In prior work, using the average of multiple PPG traces is a common practice (e.g.,~\cite{temko_accurate_2017}).
Naturally, this reduces the impact of single-channel artifacts but the output signal is still affected -- particularly when the artifacts' amplitudes are much greater than the actual signal. 
In an effort to combine PPG traces while taking their signal quality into account, Warren et al. switch between channels every two seconds to select the most promising for HR detection~\cite{warren_improving_2016}.
However, this method may lose information from channels that are not currently assumed to have the best signal quality which weighs especially heavily when the quality estimation is less-than-optimal.

\begin{figure}[t]
    \centering
    \includegraphics[width=\columnwidth]{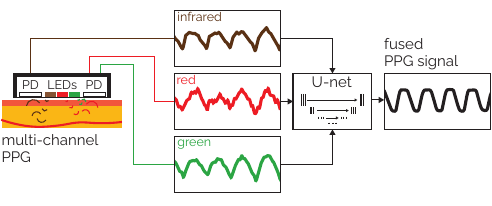}
    \caption{Our learning-based method recovers a single PPG signal from multi-wavelength input signals, thereby restoring signal morphology across them to produce an optimized output signal for subsequent assessment of cardiac dynamics, such as heart rate.}
    \label{fig:figure1}
\end{figure}

In this paper, we introduce a novel method that dynamically fuses all reflections from a multi-channel PPG signal to recover a single and robust PPG estimate signal.
As shown in Figure~\ref{fig:figure1}, our method Tri-Spectral PPG integrates a U-Net convolutional neural network to infer the most likely fused PPG sequence from all input observations.
For training and cross-validated evaluation, we introduce a novel dataset of 10~participants who wore a 3-channel reflective PPG sensor and Lead\,I electrocardiogram (ECG) digitizer for 13~hours during a variety of everyday and motion-afflicted activities outside the laboratory.

Based on the PPG signals recovered with our method, HR detection had a 23\% lower error than when computed based on green PPG reflections, which were best among all individual wavelengths.

\section{Method}

\subsection{Dataset Recording}

For our learning-based method, we recorded a novel dataset as follows. 
We recruited 10 healthy adults, comprising 7 males and 3 females, aged 22 to 69 years (mean age 37). 
The participants' skin tones ranged from 1 to 3 on the Fitzpatrick scale~\cite{fitzpatrick1988validity}.
Over a 13-hour period, we recorded their cardiac activity using optical and biopotential sensors.

\begin{figure}[b]
    \centering
    \includegraphics[width=\columnwidth]{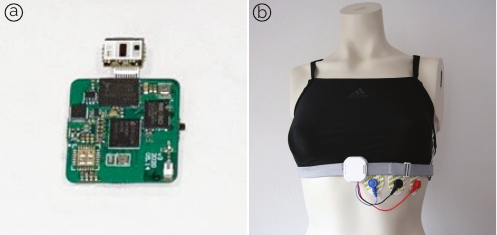}
    \caption{For evaluating our method, we captured a dataset of 10 participants who wore a standalone sensing device for 13\,hours.
    (a)~Reflective PPG was recorded at three wavelengths (green, red, infrared) using an SFH7072 module and MAX86141.
    (b)~The device additionally recorded the Lead~I ECG (MAX30003) for calculating reference metrics.}
    \label{fig:experiment}
\end{figure}

\subsubsection{Apparatus}

Each participant wore a standalone device on their sternum that integrated reflective PPG sensors as shown in Figure~\ref{fig:experiment}.
PPG measurements were obtained using an optical analog front-end at 128\,Hz (MAX86141, Analog Devices) that connected to an optical module (SFH7072, ams-OSRAM) with a green (530\,nm), a red (655\,nm), and an infrared (940\,nm) LED.
The module also included two photodiodes: an infrared-cut photodiode (402--694\,nm) was used in combination with the green and red LED, and a broadband photodiode (410--1100\,nm) was used in combination with the infrared LED.
Sensor data was continuously read by a System-on-a-Chip (DA14695, Dialog Semi),  stored in NAND memory (TH58CYG3S0HRAIJ, Kioxia Corp.), and downloaded from the device after the recording procedure.
The device was powered by a CR2032 coin cell battery.

For ground-truth cardiac activity, the device additionally collected a Lead~I ECG through a biopotential sensor (MAX30003, Analog Devices), connected to gel electrodes placed on a person's chest.
The PPG and ECG measurements were synchronized by an external clock which was generated based on the real-time clock of the System-on-a-Chip.

\subsubsection{Experimental Protocol}

Participants gathered in downtown Zurich in the morning, where an experimenter outfitted each participant with a sensor device using a chest strap.
The experimenter then ensured PPG and ECG signal quality (20\,min).
Participants then took a minivan from Zurich to Grindelwald (140\,min), transitioned to a cablecar and train to Jungfraujoch railway station at 3460\,m above sea level (80\,min), walked through the museum and exhibition area (60\,min), walked the stairs up to the observatory (60\,min), sat down for lunch (60\,min), walked through the outside area (60\,min), rested inside (60\,min), took the train and cablecar back to Grindelwald (80\,min), and returned to Zurich on the minivan (140\,min).
Finally, the experimenter removed and collected all devices from the participants (20\,min).

\subsubsection{Dataset Summary}

Across all 10 participants, we captured $\sim$13 hours of signals from the three PPG channels at 128\,Hz.
In addition, we captured the continuous Lead~I ECG per participant for the same duration, synchronized to the recorded PPG signals.
In total, our dataset thus comprises 130\,hours of recordings.

\subsection{Tri-Spectral PPG Signal Fusion Method}

We propose a learning-based signal fusion method that takes synchronized multi-wavelength PPG signals as input and recovers an optimized signal, thereby restoring signal morphology while removing artifacts.

\subsubsection{Synthesizing a PPG Reference Signal from Recordings}

\begin{figure}[t]
    \centering
    \includegraphics[width=\columnwidth]{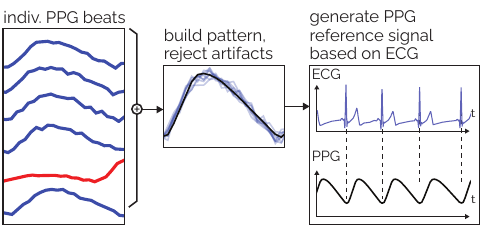}
    \caption{Synthesizing a PPG reference signal to train our learning-based method. 
    A PPG signal pattern is repeated based on ECG R-peaks, generated in windows of 5~minutes and consisting of a normalized aggregate of the input PPG waves, which are correlated with a leaning triangle wave to reject distorted and noisy input signals.}
    \label{fig:template}
\end{figure}

To train our model, we synthesized an optimized PPG signal that was true to the recorded morphology through aggregated individual PPG waves.
These patterns are resampled and concatenated to align with the R-R intervals of the reference ECG signal, using Pan-Tompkins~\cite{pan_real-time_1985} for R peak detection.
We compute the PPG signal patterns following the approach by Warren et~al.~\cite{warren_improving_2016} using ensemble averaging as shown in Figure\,\ref{fig:template}.
To account for changes in the signal morphology over time, we compute the pattern in windows of 5\,minutes.

For template formation, we split the input PPG signals using the R-peaks from the ECG signal, disregarding sections with lengths corresponding to a HR below 40 or above 185.
Each resulting section is down-sampled to 100 samples to account for HR variations and is then z-scored (i.e., standardized to zero mean and unit standard deviation).
The template is derived by averaging sections that sufficiently correlate with a leaning triangle wave ($r>.8$), excluding corrupted segments.

Finally, we synthesize the reference signal by concatenating one signal pattern for each R-R-interval.
The concatenated pattern is a weighted average of the two nearest neighboring computed patterns.
The weighting is inversely proportional to the distance of the R-R-interval to the corresponding pattern and thus smooth transitions between different patterns are guaranteed.
To match the HR and signal length, each concatenated pattern is resampled to match the corresponding interval.

\subsubsection{Network Architecture and Training}
\label{subsec:architecture}

For the task of signal recovery, our method adapts a U-Net architecture~\cite{Wave_U_Net}, using a three-channel PPG signal as input and generating a fused PPG signal as output.
We augment the model architecture with four up-sampling and four down-sampling layers. 
The four up-/down- sampling layers are optimized from 32 to 256 kernels, similar to previous work on estimating waveform sources~\cite{waveunet}.
For training, we used the Adam optimizer~\cite{Adam} with $\beta_1=0.9$, $\beta_2=0.999$, and a mini-batch size of 80.
The learning rate is initialized with $0.001$ and reduced by half when the validation loss stops improving for 50 consecutive epochs. 
The training is terminated when 75 successive epochs show no validation performance improvements.
The best model is chosen as the lowest L1 loss on the validation data.

\subsection{Evaluation}

To evaluate our method, we extracted 4000 evenly spaced out sections with a length of 8\,seconds from the recording of each participant, which corresponds to 8.9\,hours of measurements.
Besides cropping of measurements during the outfitting and removal of the sensors, sections with inconsistent or noisy ECG measurements were removed in this process while preserving a balanced data set.

\begin{table}[b]
\centering
\caption{Performance Metrics for Signal Morphology}
\label{tab:performance_metrics}
\begin{tabular}{lllllll}
\toprule
\textbf{Participant} & \textbf{MAE} & \textbf{RMSE} & $\rho_{ours}$ & $\rho_G$ & $\rho_R$ & $\rho_{IR}$ \\ 
\midrule
1                    & 0.356         & 0.281         & 0.868  & 0.652  & 0.481 & 0.621    \\ 
2                    & 0.520         & 0.547         & 0.740  & 0.631  & 0.476 & 0.610    \\ 
3                    & 0.434         & 0.390         & 0.800  & 0.728  & 0.634 & 0.673    \\ 
4                    & 0.517         & 0.513         & 0.755  & 0.593  & 0.453 & 0.572    \\ 
5                    & 0.390         & 0.289         & 0.857  & 0.671  & 0.501 & 0.619    \\ 
6                    & 0.358         & 0.278         & 0.874  & 0.642  & 0.498 & 0.654    \\ 
7                    & 0.346         & 0.243         & 0.890  & 0.638  & 0.487 & 0.580    \\ 
8                    & 0.312         & 0.331         & 0.912  & 0.616  & 0.476 & 0.574    \\ 
9                    & 0.480         & 0.458         & 0.775  & 0.640  & 0.435 & 0.628    \\ 
10                   & 0.390         & 0.328         & 0.843  & 0.569  & 0.437 & 0.556    \\ 
\midrule
\textbf{Mean}        & 0.410         & 0.368         & 0.831  & 0.645 & 0.487 & 0.608   \\
\bottomrule
\end{tabular}
\end{table}

\label{subsec:peak_and_hr}

To evaluate the fused PPG signal, we compare its validity for HR detection to the input signals and its waveform morphology.
This is done for each section of 8\,seconds separately.
We first apply a bandpass filter to the input (second-order Butterworth, 0.6--3.3\,Hz passband) and then detect peaks whenever a PPG signal crosses its moving average plus an offset, building on van Gent et~al.~\cite{van_gent_heartpy_2019}.
We determine the offset by minimizing the variance in the resulting peak intervals.
Peaks are removed if the intervals between neighbors would result in an HR greater than 185\,bpm.

During model training and evaluation, we implement five-fold cross-validation across ten subjects.
We test two participants in each fold. 
Additionally, one out of the remaining eight participants is used as validation data for early stopping and model selection. 
It is worth mentioning that the model is evaluated with subjects that were not used for training.

To reduce the impact of spurious peaks on HR computation, we only process interbeat intervals (IBI) that lie within at least five consecutive IBI where $\nicefrac{\min IBI}{\max IBI} > 0.51$.
This filtering step is relatively lenient and retains most PPG beats, even less accurate sections that more conservative filters would remove.
Our method explicitly preserves these segments to ensure the presence of signals in all segments which is required for the evaluation. 
Finally, HR values are estimated from the IBI.

Moreover, we calculate the mean absolute error (MAE), root mean square error (RMSE), and correlation ($\rho$) between the fused PPG signal and the reference signal to evaluate the accuracy of our proposed method in approximating the waveform morphology.

\section{Results}
\label{sec:results}

\subsection{Signal Morphology}

The PPG signal output by our method during inference exhibited a mean correlation of 0.831 with the synthesized reference PPG signal as shown in Table~\ref{tab:performance_metrics}.
This is higher than all contributing PPG channels which exhibited a mean correlation of 0.645 (green), 0.487 (red), and 0.608 (infrared).  

The MAE of the signal output compared to the reference signal lies within a range of 0.312 and 0.520.
The RMSE lies within a range of 0.243 and 0.547.

\begin{table}[t]
\centering
\caption{HR Detection Mean and Median Error (bpm)}
\label{tab:hr}
\begin{tabular}{lr@{  }rr@{  }rr@{  }rr@{  }r}
\toprule
\textbf{Part.} & \multicolumn{2}{c}{\textbf{green}} & \multicolumn{2}{c}{\textbf{red}} & \multicolumn{2}{c}{\textbf{ir}} & \multicolumn{2}{c}{\textbf{ours}}\\ 
\midrule
1  & 3.4 & 0.5 & 9.4  & 4.8  & 5.1  & 0.7 & 1.9  & 0.7 \\
2  & 6.5 & 1.3 & 17.2 & 11.5 & 11.7 & 6.0 & 5.4  & 1.5 \\
3  & 2.6 & 0.3 & 15.4 & 10.7 & 15.2 & 9.8 & 2.2  & 0.7 \\
4  & 4.3 & 1.1 & 11.1 & 7.9  & 8.1  & 5.5 & 2.0  & 0.9 \\
5  & 2.1 & 0.3 & 10.7 & 7.2  & 8.9  & 4.7 & 3.2  & 0.9 \\
6  & 8.7 & 3.7 & 15.5 & 10.2 & 13.6 & 8.9 & 5.7  & 1.6 \\
7  & 6.2 & 1.7 & 10.3 & 7.6  & 8.8  & 5.3 & 3.6  & 0.9 \\
8  & 6.6 & 3.4 & 15.7 & 11.2 & 12.2 & 7.9 & 10.0 & 6.6 \\
9  & 7.2 & 0.5 & 12.9 & 8.5  & 11.5 & 7.0 & 3.6  & 0.9 \\
10 & 9.0 & 4.3 & 11.6 & 8.4  & 10.3 & 6.9 & 5.2  & 1.9 \\
\midrule
\textbf{Mean}        & 5.9 & 1.9 & 13.4 & 9.2  & 11.1 & 6.9 & \textbf{4.5}  & \textbf{1.8} \\
\bottomrule
\end{tabular}
\end{table}

\subsection{HR Detection (as Downstream Task)}

When computing the HR based on the single-wavelength PPG trace, green performed best (MAE 5.9\,bpm, median absolute error 1.9\,bpm), followed by infrared (11.1\,bpm, 6.9\,bpm) and red (13.4\,bpm, 9.2\,bpm) PPG measurements as shown in \autoref{tab:hr}.
The HR based on the signal produced by our method had a MAE of 4.5\,bpm and a median absolute error of 1.8\,bpm.
This is a 23\% improvement of the MAE and a 5\% of the median absolute error over the best single channel (green).

\setlength{\tabcolsep}{4pt}

\section{Discussion and Limitations}
Our method purposely does not directly estimate metrics of cardiac activity, such as HR itself.
Instead, our approach recovers the underlying morphology of the observed and possibly noisy PPG signals and can serve as input to a variety of downstream tasks, one of which is HR as used in our evaluation.
We believe that our method can serve as input for several more applications of pulse wave analysis in the future.
Limiting factors of this work include the small sample size of 10\,Participants who all were within 1--3 on the Fitzpatrick scale, with no participants rating themselves 4--6 (darker skin tones). 
Considering the more challenging nature of PPG measurements among people with darker skin tones~\cite{fallow2013influence}, it will be important to broaden data collection in future efforts and investigate the effects of different PPG wavelengths and body locations on the results of this approach.

\section{Conclusion}
We have introduced Tri-Spectral PPG, a novel method that fuses PPG signals acquired at different wavelengths to retrieve a more robust fused signal which preserves the signal morphology and therefore allows further processing to retrieve arbitrary PPG-based measures.
Our method takes as input PPG signals recorded at different wavelengths and leverages a U-Net to produce a single output signal.
Compared to prior work, its advantage is the ability to consider information from all PPG signals, as opposed to switching between them, while also removing artifacts completely, as opposed to averaging channels.
To train our model, we have generated a synthesized reference signal using an ECG signal and an aggregation method to retrieve PPG wave patterns.
To evaluate our method, we contribute a novel dataset that we captured from 10 participants over 13 hours during outdoor and mountain activities.
The signal produced by our method correlated better with the reference signal than any contributing signal and the HR detection on the produced signal had an error that was 23--66\% lower compared to the contributing signals.
We also hope to inspire future work through the release of our model, code, and data.

\balance{}
\bibliographystyle{IEEEtran}
\bibliography{manual_refs}
\balance{}

\end{document}